\documentclass[a4paper,aps,pra,10pt,twocolumn,floatfix,superscriptaddress,notitlepage,usenames,dvipsnames,svgnames,table,nofootinbib,longbibliography]{revtex4-2}

\usepackage{tikz}
\usetikzlibrary{quantikz2}

\usepackage[margin=1in]{geometry}

\usepackage{graphicx} 
\usepackage{subcaption}
\usepackage{adjustbox}

\usepackage{amsmath}
\usepackage{braket}
\DeclareMathOperator{\Tr}{Tr}

\usepackage{ulem}

\usepackage{amssymb}
\usepackage{amsthm}
\usepackage{hyperref}
\usepackage{cleveref}

\newtheorem{theorem}{Theorem}[section]

\newtheorem{definition}{Definition}
\newtheorem{proposition}{Proposition}

\usepackage{xcolor}

\usepackage{pgfplots}
\definecolor{blue-violet}{rgb}{0.54, 0.17, 0.89}

\usepackage{hyperref}
\hypersetup{
 pdfauthor={Emanuel Dallas, Faidon Andreadakis, and Paolo Zanardi},
 pdftitle={Nonlocal Magic Generation and Information Scrambling in Noisy Clifford Circuits},
 pdflang={English},colorlinks=true,linkcolor=RubineRed,citecolor=blue-violet,urlcolor=Cerulean}

\begin{document}

\title{Nonlocal Magic Generation and Information Scrambling in Noisy Clifford Circuits}

\author{Emanuel Dallas}
\email [e-mail: ]{dallas@usc.edu}
\affiliation{Department of Physics and Astronomy, and Center for Quantum Information Science and Technology, University of Southern California, Los Angeles, California 90089-0484, USA}
\author{Paolo Zanardi}
\email [e-mail: ]{zanardi@usc.edu}
\affiliation{Department of Physics and Astronomy, and Center for Quantum Information Science and Technology, University of Southern California, Los Angeles, California 90089-0484, USA}
\affiliation{Department of Mathematics, University of Southern California, Los Angeles, California 90089-2532, USA}

\date{\today}
\begin{abstract}

In this work, we investigate the average information scrambling and nonlocal magic generation properties of random Clifford encoding-decoding circuits perturbed by local noise. We quantify these with the bipartite algebraic out-of-time order correlator ($\mathcal{A}$-OTOC) and average Pauli-entangling power (APEP) respectively. Using recent advances in the representation theory of the Clifford group, we compute both quantities' averages in the limit that the circuits become infinitely large. We observe that both display a ``butterfly effect" whereby noise occurring on finitely many qubits leads to macroscopic information scrambling and nonlocal magic generating power. Finally, we numerically study the relationship between the magic capacity \cite{capacity_robustness}, an operator-level magic monotone, of the noise channel and the APEP of the resulting circuit, which may provide insight for designing efficient nonlocal magic factories. 
\end{abstract}

\maketitle

\section{Introduction}\label{intro}

In recent decades, the field of quantum error correction (QEC) has grown tremendously. One research avenue has been motivated by the recognition that fault-tolerant quantum computation requires careful encoding of logical information in physical qubits \cite{kitaev_threshold, preskill_threshold, gottesman_qec, preskill_future_qc}. Recent theoretical work has uncovered classes of efficient quantum error-correcting codes (QECCs) \cite{asymp_ldpc, fiber_bundle_codes, balanced_product_codes, tanner_codes, good_qldpc}, and experimental results have demonstrated beyond-threshold encoding and error-correction schemes on existing hardware \cite{google_qec, quantinuum_qec, aws_qec, yale_qec}. Other research has uncovered deep connections between quantum error correcting codes and fundamental physics, such as the AdS/CFT correspondence \cite{qec_adscft, qec_adscft_2, qec_adscft_3} and topological phases of matter \cite{qec_top_1, qec_top_2, qec_top_3}.

Most work has focused on \textit{stabilizer} codes, in which logical states lie within the joint $+1$ eigenspace of a stabilizer group, an abelian subgroup of the Pauli group $\mathcal{P}_n$ \cite{gottesman_thesis}. The encoding operations that generate such codes lie within the Clifford group \cite{gottesman_thesis}, the normalizer of the Pauli group. It is well known, due to the Gottesman-Knill theorem, that Clifford circuits are efficiently simulable classically and therefore insufficient for universal quantum computation \cite{gottesman_knill}.

This realization has motivated research into the resource theory of ``non-stabilizerness," colloquially known as \textit{magic}. Magic states and/or magic-generating operations are necessary for universal quantum computation, and much work has been devoted to developing efficiently calculable magic monotones \cite{mana, robustness_of_magic, stabilizer_renyi_entropy, thauma} as well as concentrating magic from multiple sources in individual states, known as magic state distillation \cite{magic_distillation_2011, magic_distillation_experiment, magic_distillation_2023}.

Recently, several works have focused on understanding \textit{nonlocal} magic \cite{nlm_1, nlm_2, nlm_3, nlm_4, nlm_5, apep}. In the same way that in QECCs logical information is stored nonlocally to protect it against local noise, magic can be distributed nonlocally for the same purpose. Understanding and quantifying nonlocal magic is thus important for determining how to robustly and fault-tolerantly generate and store magic (see \cref{cliff_apep}).

In \cite{butterfly_effect}, we studied the information scrambling properties of a toy model of noisy quantum encoding-decoding circuits, inspired by \cite{Turkeshi}. In the present work, we investigate a similar model and study information scrambling and nonlocal magic generation induced by noisy stabilizer encodings. We consider an $L$-qubit circuit consisting of a Clifford-random unitary encoding $C$, followed by identical noise processes $\mathcal{E}$ on $k$ qubits, followed by $C^\dagger$. This allows us to compare the scrambling sensitivity of Clifford circuits to fully Haar-random ones, as well as understand how local magic operations coupled with interacting Clifford operations generate nonlocal magic.The nonlocal magic generation by this model is identical to one without the initial Clifford encoding (see \cref{omega_prime}); such a setup, in which a single-qubit magic gate $U$ is applied simultaneously to $k$ qubits before a Clifford encoding $C$, represents an even more generic way to generate nonlocal magic.

Our scrambling metric, as in \cite{butterfly_effect}, is the bipartite algebraic out-of-time-order correlator ($\mathcal{A}$-OTOC \cite{aotoc_faidon, StyliarisBipartite, NamitBipartite}), which quantifies information scrambling between two subsystems. The $A$-OTOC includes a Haar average over observables in each subsystem and thereby depends solely on the selected bipartition and circuit. Our nonlocal magic measure is the average Pauli-entangling power (APEP), introduced in \cite{apep}. This is the average operator entanglement, over the selected bipartition, of Paulis evolved by the circuit. If the bipartition is defined with respect to physical qubits, then every Pauli is a product operator and has zero operator entanglement over the bipartition. If the APEP of a circuit is non-vanishing, then some Paulis must get mapped to non-Paulis, which means the circuit is non-Clifford. The APEP is thus a ``proxy"\footnote{The monotonicity of the APEP has not been shown.} for magic distributed nonlocally over the bipartition. In this work, we select our bipartition to be between two equal-sized subsets of qubits ($\mathcal{H} \cong (\mathbb{C}^2)^{L/2} \otimes (\mathbb{C}^2)^{L/2}$), allowing us to study extensive properties of these circuits.

We derive analytic formulae for the $A$-OTOC and APEP that depend on the noise channel $\mathcal{E}$ and $L$. In the limit that $L \rightarrow \infty$, the formula dramatically simplifies. From this, we immediately observe a butterfly effect in which noise on finitely many qubits causes ``macroscopic" scrambling and non-logical magic generation. We then numerically investigate the relationship between the \textit{magic capacity} of noisy unitaries and the APEP of the corresponding circuits. We discuss the practical trade-offs that are entailed by this relationship.

In \cref{mathprelim}, we provide mathematical background on the $A$-OTOC and the APEP. In \cref{physsetup}, we introduce the physical circuit model we will analyze. In \cref{mainresults}, we present formulae for the $A$-OTOC and APEP under various limits and noise channels, and derive an expression for it in the $L \rightarrow \infty$ limit. In \cref{examples}, we present two examples that analytically demonstrate the butterfly effect. In \cref{magcapandapep}, we compare a measure of magic for single-qubit coherent noise unitaries with the resultant noisy circuit APEP to connect local magic injection with nonlocal magic propagation. Finally, \cref{conclusion} provides a summary and avenues for future work.

\section{Mathematical preliminaries}\label{mathprelim}

\subsection{The general \texorpdfstring{$\mathcal{A}$-OTOC}{A-OTOC}}

Let $\mathcal{H} \cong \mathbb{C}^d$ be a finite-dimensional Hilbert space representing a quantum system. Any physical observable within this system is described by a linear operator, and we denote the space of all such operators by $\mathcal{L}(\mathcal{H})$. This space, $\mathcal{L}(\mathcal{H})$, also forms a Hilbert space, equipped with the Hilbert-Schmidt inner product, defined as $\langle X, Y\rangle=\Tr\left(X^{\dagger} Y\right)$ for operators $X$ and $Y$. In the Heisenberg picture, the evolution of a physical observable $X \in \mathcal{L}(\mathcal{H})$ in an open quantum system is given by $\mathcal{E}(X)$, where $\mathcal{E}^\dagger$ is the CPTP map governing state evolution.

The key mathematical objects in this discussion are hermitian-closed, unital subalgebras $\mathcal{A} \subset \mathcal{L}(\mathcal{H})$, which describe the degrees of freedom of interest. The commutant algebra $\mathcal{A}^{\prime}=\left\{Y \in \mathcal{A}^{\prime} \mid[Y, X]=0 \text{ for all } X \in \mathcal{A}\right\}$ captures the symmetries of $\mathcal{A}$ and corresponds to degrees of freedom that are initially uncorrelated with $\mathcal{A}$. By the double commutant theorem \cite{doublecommutantthm}, $\left(\mathcal{A}^{\prime}\right)^{\prime}=\mathcal{A}$, so these algebras can be viewed as pairs $\left(\mathcal{A}, \mathcal{A}^{\prime}\right)$. During time evolution, information is exchanged between $\mathcal{A}$ and $\mathcal{A}^{\prime}$, which we quantify with the $\mathcal{A}$-OTOC \cite{aotoc_faidon}.

\begin{definition}\label{aotoc_def}
    The \textbf{$\mathcal{A}$-OTOC} of algebra $\mathcal{A}$ and channel $\mathcal{E}$ is defined as:
    \begin{equation}
        G_{\mathcal{A}}\left(\mathcal{E}\right)=\frac{1}{2 d} \mathbb{E}_{X_{\mathcal{A}}, Y_{\mathcal{A}^{\prime}}}\left[\left\|\left[X_{\mathcal{A}}, \mathcal{E}\left(Y_{\mathcal{A}^{\prime}}\right)\right]\right\|_2^2\right]
    \end{equation}
$\mathbb{E}_{X_{\mathcal{A}}, Y_{\mathcal{A}^{\prime}}}$ denotes the Haar average over the unitaries $X_{\mathcal{A}} \in \mathcal{A}$ and $Y_{\mathcal{A}^{\prime}} \in \mathcal{A}^{\prime}$. 
\end{definition}

The evolution of operators in $\mathcal{A}^{\prime}$ under $\mathcal{E}$ may lead to non-commutativity with operators in $\mathcal{A}$, which is interpreted as scrambling between the corresponding degrees of freedom.

In this paper, we reserve $\mathcal{E}$ to refer to the map in the Heisenberg picture, and we consider maps that are CPTP in the Schrodinger picture (i.e. $\mathcal{E}^\dagger$ is CPTP). This implies that $\mathcal{E}$ is unital ($\mathcal{E}(\mathbb{I}) = \mathbb{I}$).

\subsection{Bipartite \texorpdfstring{$\mathcal{A}$-OTOC}{A-OTOC}}

We now consider the case of bipartite algebras. Let the Hilbert space $\mathcal{H} \equiv \mathcal{H}_A \otimes \mathcal{H}_B$. Our algebra is now $\mathcal{A} = \mathbb{I}_A\otimes L(\mathcal{H}_B)$, with $\mathcal{A}' = L(\mathcal{H}_A)\otimes \mathbb{I}_B$. Henceforth, we will denote the bipartite $\mathcal{A}$-OTOC as $G$.

In \cite{NamitBipartite}, the following was shown: 
\begin{proposition}\label{aotoc_swap}
    Let $S \equiv S_{A A^{\prime} B B^{\prime}}$ be the swap operator over $\mathcal{H}_{A B} \otimes \mathcal{H}_{A^{\prime} B^{\prime}}$, and let $S_{AA'}$ be the swap over $\mathcal{H}_A$ and $\mathcal{H}_{A'}$ on the same space $\mathcal{H}_A \otimes \mathcal{H}_{B} \otimes \mathcal{H}_{A'} \otimes \mathcal{H}_{B'}$. Then for a quantum channel $\mathcal{E}: \mathcal{L}\left(\mathcal{H}_{A B}\right) \rightarrow$ $\mathcal{L}\left(\mathcal{H}_{A B}\right)$, the bipartite $\mathcal{A}$-OTOC is:
    \begin{equation}
        G(\mathcal{E})=\frac{1}{d^2} \operatorname{Tr}\left(\left(d_B S-S_{A A^{\prime}}\right) \mathcal{E}^{\otimes 2}\left(S_{A A^{\prime}}\right)\right).
    \end{equation}
\end{proposition}

We use this formula to derive our results in \cref{mainresults}.

\subsection{Average Pauli-entangling Power}

We begin by defining operator entanglement over a bipartition $\mathcal{H} \cong \mathcal{H}_A \otimes \mathcal{H}_B$. Let $O$ be an arbitrary operator in $\mathcal{U}_n$, the unitary group over $n$ qubits. Then $O$ has the Schmidt decomposition $O/\sqrt{d} = \sum_i \sqrt{\lambda_i} V_i \otimes W_i$, where $\{V_i\}$ and $\{W_i\}$ are orthonormal sets on $\mathcal{H}_A$ and $\mathcal{H}_B$ respectively, with respect to the Hilbert-Schmidt inner product, and $\{\lambda_i \in \mathbb{R}^+\}$ forms a probability distribution, $\sum_i \lambda_i = 1$. We define:

\begin{definition}[Linear Operator Entanglement]
    $E_{\text{lin}}(O) = 1 - \sum_i \lambda_i^2$.
\end{definition}

We work with the above linear entropy for the sake of analytic tractability. Given that we are interested in a bipartition over physical qubits, i.e. $\mathcal{H} \cong (\mathbb{C}^2)^{\otimes N_A} \otimes (\mathbb{C}^2)^{\otimes N_B}$, every Pauli $P \in \mathcal{P}_n$ is a product operator over this bipartition and therefore has $E_{lin}(P) = 0$. As Cliffords map Paulis to Paulis, then $E_{lin}(CPC^\dagger) = 0 \ \forall C \in \mathcal{C}_n$. This motivates the study of the growth of $E_{lin}$ of Paulis under an evolution as a measure of the evolution's non-stabilizerness over a bipartition. Our operator-level nonlocal magic ``proxy" is the:

\begin{definition}[Average Pauli-entangling Power (APEP)]
    Let $\tilde{\mathcal{P}}_N \equiv \mathcal{P}_N/\mathbb{Z}_4$ be the abelianized Pauli group (modulo phases). Then the APEP \begin{equation}
        P_E(U) \equiv \mathbb{E}_{P \in \tilde{\mathcal{P}}_N} E_{lin}(U^\dagger PU).
    \end{equation}
\end{definition}

We point out two results from \cite{apep}. First, the APEP is invariant under pre-processing with Clifford unitaries -- $P_E(U) = P_E(UC) \ \forall C \in \mathcal{C}_n$. Second, consider $\mathcal{H}^{\otimes 4} \cong (\mathcal{H}_A \otimes \mathcal{H}_B)^{\otimes 4}$. Then

\begin{equation}\label{APEP_eqn}
    P_E(U) = 1 - \frac{1}{d^2} \, \Tr \left( T^{A}_{(12)(34)} \, U^{\dagger \otimes 4} Q U^{\otimes 4} \right),
\end{equation}

where $T^{A}_{(12)(34)}$ is a representation of the $(12)(34)$ permutation acting on the four copies of the $A$ subsystem, and $Q \equiv \frac{1}{d^2}\sum_{P \in \tilde{\mathcal{P}}_N} P^{\otimes 4}$. We note that, up to a reordering of tensor factors, $Q = \frac{1}{d^2} \Lambda^{\otimes N}$, where $\Lambda \equiv \sum_{P \in \tilde{\mathcal{P}}_1} P^{\otimes 4}$, the sum over four copies of the standard single-qubit Paulis; $Q$ is an essential ingredient of Clifford group representation theory \cite{cliff_gracefully} and $\Lambda$ appears in our own results. We use \cref{APEP_eqn} to derive \cref{cliff_apep}.

\section{Encoding-Decoding Circuit Model}\label{physsetup}

Consider a chain of $L$ qubits, where $L$ is even. 
Let $C$ be an $L$-qubit Clifford unitary, 
and let $\mathcal{E}$ be a single-qubit CP map. We denote the adjoint action of $C$, in the Heisenberg picture, as $\mathcal{C}$.

Consider the map $\Omega_{C,\mathcal{E}} \equiv \mathcal{C^\dagger}(\mathcal{E}^{\otimes k} \otimes \mathbb{I}^{\otimes (L-k)})\mathcal{C}$ in the Heisenberg picture. Assume, without loss of generality, that $\mathcal{E}$ is applied to the ``first" $k$ qubit sites.

\begin{figure}
    \centering
    \begin{quantikz}[row sep = {0.6cm, between origins}]
    &\gate[4, style={rounded corners, fill=red!20}][1cm]{C}&[0.3cm]\measure[style = {fill=yellow}]{\mathcal{E}}\gategroup[wires=2,steps=2,style={inner xsep=0pt, inner ysep=0pt, outer sep=0pt, draw=none}, label style={label position=right, anchor=east, xshift=-0.75cm, yshift = -0.2cm}]{\parbox{3cm}{\baselineskip=0.4\baselineskip k qubits \\ \makebox[1.3cm]{.} \\ \makebox[1.3cm]{.} \\\makebox[1.3cm]{.}}}
    &[0.5cm]\gate[4, style={rounded corners, fill=red!20}][1cm]{C^\dagger}&\\[1.2cm]
    &&\measure[style = {fill=yellow}]{\mathcal{E}}&& \\
    &&&& \\
    &&&& 
    \end{quantikz}

    \caption{Circuit diagram of $\Omega$.}
\end{figure}
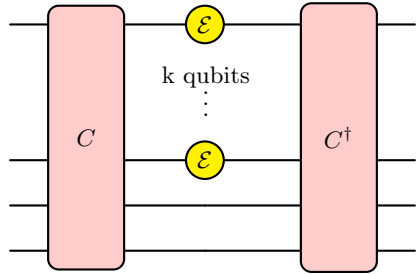

The map $\Omega_{C,\mathcal{E}}$ corresponds to a unitary evolution, a layer of single-qubit perturbations, and a reversal of the unitary evolution. The extensive properties of this map elucidate the extent to which local, possibly infinitesimal (in system size) perturbations can cause macroscopic effects. 

In what follows, we compute the average $\mathcal{A}$-OTOC over the uniform measure on Cliffords $C$, denoted as $\overline{G(\Omega_{C,\mathcal{E}})} ^C\equiv \mathbb{E}_C G(\mathcal{C}^\dagger (\mathcal{E}^{\otimes k} \otimes \mathbb{I}^{\otimes (L-k)}) \mathcal{C})$.

\section{Main results}\label{mainresults}

We are interested in taking averages over the Clifford group of both the bipartite $\mathcal{A}$-OTOC for generic single-qubit noise and the APEP for coherent noise.

To arrive at results for this model, we insert $\Omega$ in place of the generic $\mathcal{E}$ in \cref{aotoc_swap}. This expression contains four copies of the Clifford channel $\mathcal{C}$. Taking this average requires the use of Weingarten calculus \cite{weingarten_1} for the fourth moment of the Clifford group, which is significantly more complicated than for the full unitary group. Fortunately, efficient analytical constructions have recently been derived \cite{cliff_comm} and implemented \cite{dowling_cliff}.

We analytically compute this average by linearizing the expression using the swap trick \cite{mele_haar} and then employing Schur-Weyl duality \cite{weingarten_1}. The full details are presented in \cref{clifford_schurweyl}. This average involves a large number of terms and group-theoretic data as inputs, so the calculation is implemented in a Mathematica notebook.

In the following results, we use:

\begin{definition}[Natural representation]\label{naturalrep}
    Let $\mathcal{E}$ have the Kraus representation:
\begin{equation}
    \mathcal{E} = \sum_i K_i (\cdot) K_i^\dagger.
\end{equation}

Then $X \equiv \sum_i K_i \otimes K_i^\dagger$ is the \textbf{natural representation} of $\mathcal{E}$.
\end{definition}

A given map has a unique natural representation, i.e. it is independent of any particular choice of Kraus operators \cite{naturalrepresentation}. Since we work only with generic $\mathcal{E}$, we omit any subscript label on $X$ that explicitly refers to $\mathcal{E}$.

\begin{theorem}[$\overline{G(\Omega_{C,\mathcal{E}})}^C$ for infinite qubit chains]\label{cliff_aotoc}
Let $\mathcal{E}$ be a CPTP map with natural representation $X$. In the limit that $L \rightarrow \infty$, the Clifford-averaged symmetric bipartite $\mathcal{A}$-OTOC is given by:

\begin{equation}
    \overline{G(\Omega)}^C = \left|\left|\frac{X}{2}\right|\right|_2^{2k} - \Tr^k(S(X \otimes X^{\dagger})\Lambda).
\end{equation}
\end{theorem}

For comparison, we restate the main result of \cite{butterfly_effect}:
\begin{theorem}[$\overline{G(\Omega_{U,\mathcal{E}})}^{Haar}$ for infinite qubit chains]\label{maintheorem} In the limit that $L \rightarrow \infty$, the Haar-averaged symmetric bipartite $A$-OTOC is given by:
\begin{equation}
    \overline{G(\Omega_{U,\mathcal{E}})}^{Haar} = \left|\left|\frac{X}{2}\right|\right|_2^{2k} - \left(\frac{\Tr(X)}{4} \right)^{2k}.
\end{equation}
\end{theorem}

We explicitly compare these formulae in the examples below. As in the Haar case, we see a butterfly effect in the Clifford average of the bipartite $\mathcal{A}$-OTOC. Cliffords have a similar tendency to scramble information, though we will see explicitly that they do not tend to scramble as much as Haar.

We use the same methods to calculate the APEP. We now restrict the noise channel $\mathcal{E} = \mathcal{U}$ to be unitary, as the APEP is only defined for unitary maps; now, $\Omega_{C,\mathcal{E}} = \Omega_{C,\mathcal{U}} = \mathcal{C}^\dagger (\mathcal{U}^{\otimes k} \otimes \mathbb{I}^{\otimes (L-k)}) \mathcal{C}$. We take $T_\sigma$ to be the representation of the permutation $\sigma \in S_4$ on the 4-tensor copy space of $L$ qubits, and we write permutations by their cyclic form in the following.

\begin{figure}
    \centering
    \begin{quantikz}[row sep = {0.6cm, between origins}]
    &&[0.3cm]\measure[style = {fill=yellow}]{\mathcal{E}}\gategroup[wires=2,steps=2,style={inner xsep=0pt, inner ysep=0pt, outer sep=0pt, draw=none}, label style={label position=right, anchor=east, xshift=-0.75cm, yshift = -0.2cm}]{\parbox{3cm}{\baselineskip=0.4\baselineskip k qubits \\ \makebox[1.3cm]{.} \\ \makebox[1.3cm]{.} \\\makebox[1.3cm]{.}}}
    &[0.5cm]\gate[4, style={rounded corners, fill=red!20}][1cm]{C^\dagger}&\\[1.2cm]
    &&\measure[style = {fill=yellow}]{\mathcal{E}}&& \\
    &&&& \\
    &&&& 
    \end{quantikz}

    \caption{Circuit diagram of $\Omega'$. $P_E(\Omega') = P_E(\Omega)$.}\label{omega_prime}
\end{figure}
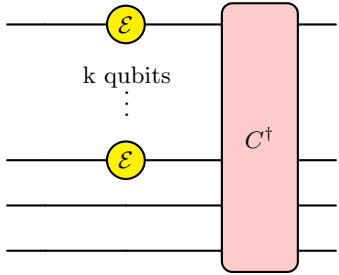

\begin{theorem}[ $\overline{P_E(\Omega_{C,\mathcal{U}})}^C$ for infinite qubit chains]\label{cliff_apep}

In the limit that $L \rightarrow \infty$, the Clifford-averaged APEP is given by:

\begin{equation}
    \begin{aligned}
        \overline{P_E(\Omega_{C,\mathcal{U}})}^C 
        &
        = 1- \quad \frac{1}{4}
        \left(
        \Tr^k(T_{(1234)}U^{\otimes 4} \Lambda (U^{\dagger})^{\otimes   4} \Lambda) + \right.
        \\
        & \quad
        + 2\Tr^k(T_{(12)(34)}U^{\otimes 4} \Lambda (U^{\dagger})^{\otimes 4} \Lambda) +
        \\
        & \left. \quad
        + \Tr^k(T_{(13)(24)}U^{\otimes 4} \Lambda (U^{\dagger})^{\otimes 4} \Lambda)
        \right).
    \end{aligned}    
\end{equation}
\end{theorem}

We recall that $P_E$ is invariant under pre-processing with Clifford unitaries. Recall that $\Omega_{C,\mathcal{U}} = \mathcal{C}^\dagger (\mathcal{U}^{\otimes k} \otimes \mathbb{I}^{\otimes (L-k)}) \mathcal{C}$. Let $\Omega'_{C,\mathcal{U}} = \mathcal{C}^\dagger (\mathcal{U}^{\otimes k} \otimes \mathbb{I}^{\otimes (L-k)})$ (see \cref{omega_prime}). Then $P_E(\Omega_{C,\mathcal{U}}) = P_E(\Omega'_{C,\mathcal{U}})$ and thus $\overline{P_E(\Omega_{C,\mathcal{U}})}^C = \overline{P_E(\Omega'_{C,\mathcal{U}})}^C$.

While $\Omega_{C,\mathcal{U}}$ represents a noisy Clifford encoding-decoding circuit, we can think of $\Omega'_{C,\mathcal{U}}$ as representing a deliberate way to locally inject magic via single-qubit unitaries before dispersing it nonlocally with a Clifford circuit. This gives an operational meaning to \cref{cliff_apep} which we discuss further in \cref{magcapandapep}.

\section{Examples}\label{examples}

\subsection{Unitary noise (Bloch sphere rotation)}

Any single-qubit unitary can be represented by a rotation of the Bloch sphere by an angle $\theta$ about some axis $\hat{n}$. For Haar-random circuits, the axis choice does not affect results, since Haar-random unitaries are not defined with respect to a particular basis. However, Cliffords are defined with respect to the Pauli basis and the unitary rotation axis affects the scrambling properties of $\Omega$.

We parameterize an arbitrary single-qubit unitary $U$ with the coordinates $\{\theta, \gamma, \phi \}$, such that $U$ is a rotation of angle $\theta$ around the Bloch vector $\vec{n} = \left(\sin{\gamma}\cos{\phi}, \sin{\gamma}\sin{\phi}, \cos{\gamma}  \right)$, i.e. $U = e^{-i \frac{\theta}{2}\vec{n}\cdot \vec{\sigma}}$. This gives\footnote{The simplification of the APEP expression for arbitrary unitaries is computationally infeasible.}:


\begin{multline}\label{eq:G-avg}
    \overline{G(\Omega_{C, \mathcal{U}})}^C = \\
    \shoveleft{1 -} \\
    \shoveleft{\left(\frac{3 + \cos 2\theta - 8\sin^2 \gamma \left(\cos^2 \gamma
    + \sin^2 \gamma \sin^2 \phi \cos^2 \phi \right)\sin^4 \tfrac{\theta}{2}}{4}\right)^k.}
\end{multline}

It can be derived that $\overline{G(\Omega_{C, \mathcal{U}})}^C$ is maximized at a rotation of $\theta = \frac{2\pi}{3}$ about the Bloch axis $\vec{n} = \left(\pm \frac{1}{3}, \pm \frac{1}{3}, \pm \frac{1}{3} \right)$, the maximally unbiased axis with respect to the Paulis. In this case, $\overline{G(\Omega_{C, U})}^C = 1 - \left( \frac{1}{4}\right)^k$.

We now explicitly consider the case of rotations about the $Z$ axis -- by symmetry, the following results hold for rotations about any Pauli axis on the Bloch sphere. Let $U = e^{-i Z\cdot\theta/2}$, and let $\mathcal{E} \equiv \mathcal{U}$ act on $k$ qubits. Then, via \cref{cliff_aotoc} and \cref{cliff_apep} respectively,
\begin{equation}\label{aotoc_unitary_clifford}
\overline{G(\Omega_{C,\mathcal{U}})}^C = 1-\frac{1}{4^k}\left(3+\cos(2\theta) \right)^k,
\end{equation}

and

\begin{equation}\label{apep_unitary_clifford}
    \overline{P_E(\Omega_{C,\mathcal{U}})}^C = 1-\frac{1}{8^k}\left(7+\cos(4\theta) \right)^k.
\end{equation}

\begin{figure*}[t!]
    \centering
    \begin{subfigure}[t]{0.48\textwidth}
    \includegraphics[width=0.6\linewidth]{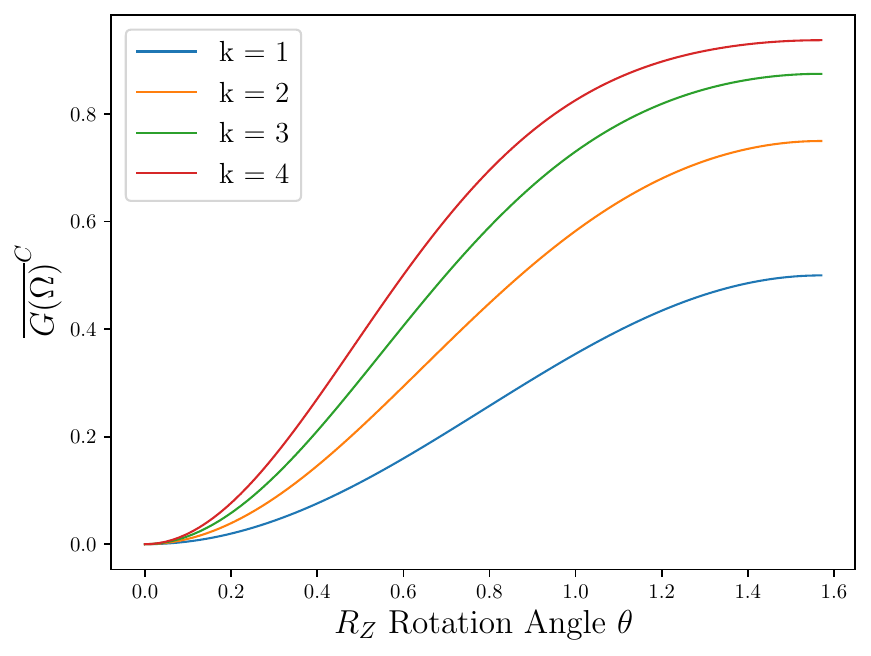}
    \caption{$\overline{G(\Omega_{C,\mathcal{U}})}^C$ vs. $\theta$ for small values of $k$.}
    \end{subfigure}%
    ~
    \begin{subfigure}[t]{0.48\textwidth}
    \includegraphics[width=0.6\linewidth]{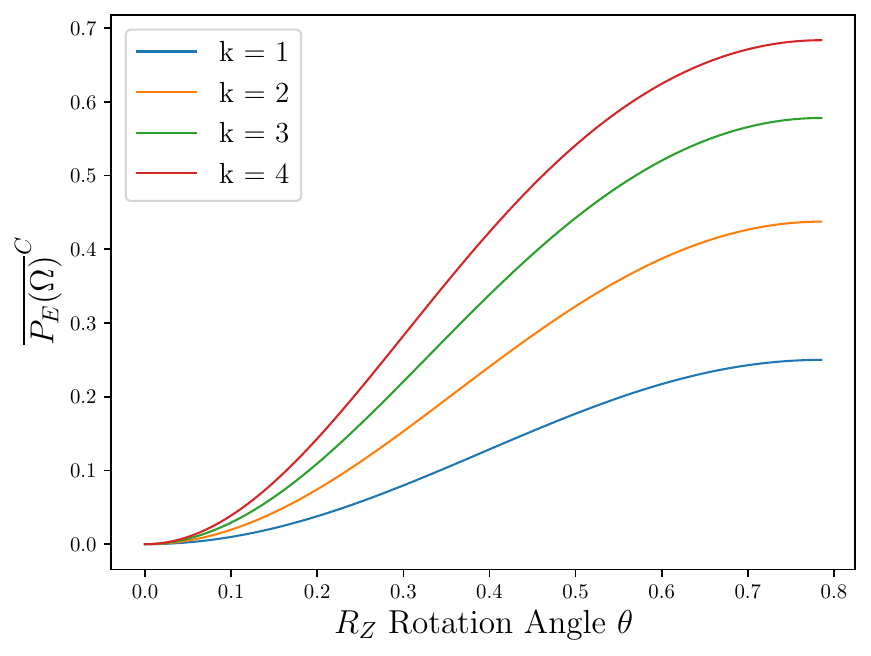}
    \caption{$\overline{P_E(\Omega_{C,\mathcal{U}})}^C$ vs. $\theta$ for small values of $k$.}
    \end{subfigure}
    \caption{$\overline{G(\Omega_{C,\mathcal{U}})}^C$ and $\overline{P_E(\Omega_{C,\mathcal{U}})}^C$ vs. $R_Z$ rotation angle $\theta$ for $k=1$ to $k=4$. As $k$ increases, both quantities increase.}
\end{figure*}

Unlike the Haar-random $\mathcal{A}$-OTOC under unitary noise \cite{butterfly_effect}, both of these quantities \textit{cannot} achieve their maxima for all values of $k$. In the Haar-random case, even $k = 1$ could give a maximal $\overline{G(\Omega)}^{Haar} = 1$ for $\theta = \pi/2$. In the Clifford case, at $\theta = \pi/2$, $\overline{G(\Omega_{C,\mathcal{U}})}^C = 1-(\frac{1}{2})^k$. This goes to $1$ only in the limit that $k \rightarrow \infty$. Likewise, $\overline{P_E(\Omega_{C,\mathcal{U}})}^C = 1-(\frac{3}{4})^k$ goes to $1$ only in the limit that $k \rightarrow \infty$.

While there is a similar butterfly effect evidenced by non-vanishing values of both quantities in the case of finite $k$, we have a fundamental distinction between the scrambling and nonlocal magic generation properties of random Clifford circuits of finitely many and infinitely many noisy qubits, a distinction not present in the Haar-random case. Intuitively, the additional scrambling ``power" that Haar-random unitaries have beyond Clifford-random unitaries can only be compensated for by extensively scaling quantities of noisy individual qubits.

\subsection{Depolarization}

Consider the $\mathcal{E}$ that fully depolarizes with probability $p$ and acts as identity with probability $1-p$; for an operator $O$, $\mathcal{E}(O) = p\mathbb{I} + (1-p)O$. One Kraus decomposition for $\mathcal{E}$ is $\left\{\sqrt{1-\frac{3p}{4}}\mathbb{I}, \frac{\sqrt{p}}{2}\sigma^x,\frac{\sqrt{p}}{2}\sigma^y, \frac{\sqrt{p}}{2}\sigma^z \right\}$.

We first recall the result for Haar-random circuits undergoing depolarization on $k$ qubits: \begin{equation}
    \overline{G(\Omega_{U,\mathcal{E}})}^{Haar} = \left( 1 - \frac{3p}{2} + \frac{3p^2}{4}\right)^k - \left(1-\frac{3p}{4}\right)^{2k}.
\end{equation}

For any finite $k$ and non-zero $p$, $\overline{G(\Omega_{U,\mathcal{E}})}^{Haar}$ has a finite value -- depolarization induces macroscopic scrambling in Haar-random circuits.

In marked contrast, for all $p$ and $k$, 
\begin{equation}
    \overline{G(\Omega_{C,\mathcal{E}})}^C = 0. 
\end{equation}

\begin{proof}
We show that this holds for all $\Omega_{C,\mathcal{E}} = \mathcal{C}^{\dagger}(\mathcal{E}^{\otimes k} \otimes \mathbb{I}^{\otimes (L-k)})\mathcal{C}$. From \cref{aotoc_def}, it is sufficient to show that $\left[X_{\mathcal{A}}, \Omega\left(Y_{\mathcal{B}}\right)\right] = 0 \  \forall X_A, Y_B$. To prove this statement, since the Paulis form a complete basis for $L(\mathcal{H})$, it is sufficient to show that $\left[P_{\mathcal{A}}, \Omega_{C,\mathcal{E}}\left(P_{\mathcal{B}}\right)\right] = 0 \  \forall P_A, P_B \in \mathcal{P}_{L/2}$.

Now, we replace $\Omega_{C,\mathcal{E}}$ with:


\begin{equation}
\begin{aligned}
    \left[P_{\mathcal{A}}, \mathcal{C}^{\dagger}(\mathcal{E}^{\otimes k} \otimes \mathbb{I}^{\otimes (L-k)})\mathcal{C}\left(P_{\mathcal{B}}\right)\right] 
    &= \\
    \left[\mathcal{C}(P_{\mathcal{A}}), (\mathcal{E}^{\otimes k}\otimes \mathbb{I}^{\otimes (L-k)})\mathcal{C}\left(P_{\mathcal{B}}\right)\right].
\end{aligned}
\end{equation}

We observe that $\mathcal{C}(P_{\mathcal{A}})$ and $\mathcal{C}\left(P_{\mathcal{B}}\right)$ are both Pauli strings. Moreover, since $\left[P_{\mathcal{A}}, P_{\mathcal{B}} \right] = 0$ by assumption, it follows by unitarity that $\left[\mathcal{C}(P_{\mathcal{A}}), \mathcal{C}(P_{\mathcal{B}}) \right] = 0$.

Now, consider the effect of $\mathcal{E}^{\otimes k}\otimes \mathbb{I}^{\otimes (L-k)}$ on a Pauli string $P = \bigotimes_j P_j$. It is straightforward to show that, for $\mathcal{E}$ acting on the $i$th qubit, $\mathcal{E}(P) = P$ if $P_i = \mathbb{I}$ and $(1-p)P$ if $P_i \neq \mathbb{I}$. From this, $(\mathcal{E}^{\otimes k}\otimes \mathbb{I}^{\otimes (L-k)})(P) = (1-p)^n P$, where $n$ is the number of depolarization sites on which $P_j \neq \mathbb{I}$. This implies that 
\begin{equation}
\begin{aligned}
    \left[\mathcal{C}(P_{\mathcal{A}}), (\mathcal{E}^{\otimes k}\otimes \mathbb{I}^{\otimes (L-k)})
        \mathcal{C}\left(P_{\mathcal{B}}\right)\right] = \\
    (1-p)^n \left[\mathcal{C}(P_{\mathcal{A}}), \mathcal{C}(P_{\mathcal{B}})\right] = 0
\end{aligned}
\end{equation}

which completes the proof.

\end{proof}

In this case of unitary noise, Clifford-random circuits displayed greater ``resilience" to noise-induced information scrambling. In the case of depolarization, they display no scrambling. 

\section{Coherent Noise Magic Capacity and Circuit Average Pauli-Entangling Power}\label{magcapandapep}

Here, we consider the relationship between the magic of coherent error operators and the APEP. Namely, how much magic resource must be ``spent" in local operations to generate nonlocal magic in Clifford encodings? To investigate this, we numerically analyze the relationship between the \textit{magic capacity}, a magic monotone introduced for CPTP maps in \cite{capacity_robustness}, of the noisy circuit's single-qubit noise operators and the Clifford-averaged APEP of the corresponding circuit.

To define magic capacity, we must first define the \textit{robustness of magic} for states \cite{capacity_robustness}. Let STAB$_n$ denote the set $n$-qubit stabilizer states. Note that the cardinality of this set is in general greater than $2^n$ \cite{stab_cardinality} and the pure states of STAB$_n$ are an overcomplete basis for the full space of $n$-qubit states. Therefore, any state $\rho$ can be written \textit{non-uniquely} as: $\rho = \sum_i q_i \ket{\phi_i}\bra{\phi_i}$, where $\sum_i q_i = 1$ for normalization. Critically, though, the $q_i$ need not all be positive; this suggests the following, shown to be a magic monotone in \cite{capacity_robustness}.


\begin{definition}\label{robustnessofmagic}
The robustness of magic is defined as:
\begin{multline}
    \mathcal{R}(\rho) =\\
    \shoveleft{\min_{\vec{q}}\Big\{ \|\vec{q}\|_{1} :
    \sum_{i} q_{i}\left|\phi_{i}\right\rangle\!\left\langle\phi_{i}\right| = \rho,\ 
    \left|\phi_{i}\right\rangle \in \operatorname{STAB}_{n} \Big\}.}
\end{multline}
\end{definition}

Intuitively, the robustness of magic captures a "distance" a state $\rho$ lies from the stabilizer polytope $\left\{\sigma: \sum_{i} s_{i}\left|\phi_{i}\right\rangle\left\langle\phi_{i}\right|=\sigma, \ s_i \geq 0 \ \forall i, \ \left|\phi_{i}\right\rangle \in \operatorname{STAB}_{n}\right\}$.

Now, for an $n$-qubit CPTP map $\mathcal{M}$, we define the:

\begin{definition}[Magic Capacity]\label{magiccapacity}
    \begin{equation}
    \mathcal{K}(\mathcal{M})=\max _{|\phi\rangle \in \operatorname{STAB}_{2 n}} \mathcal{R}\left[\left(\mathcal{M} \otimes \mathbb{I}_{n}\right)|\phi\rangle\langle\phi|\right].
\end{equation}
\end{definition}

It is generally difficult to arrive at analytic formulae for the magic capacity of particular operators. We thus numerically compute the magic capacity and plot it against the average APEP. The data look strikingly similar to the results in \cref{aotoc_unitary_clifford} and \cref{apep_unitary_clifford}. We therefore used $\overline{P_E(\Omega_{C,\mathcal{U}})} = 1-\left\vert\cos(a\cdot(\mathcal{K}(\Omega_{C,\mathcal{U}})-1))\right\vert^{b\cdot k}$ as an ansatz with which to jointly fit the data for $k$ values from $1$ to $20$. Additional details and plots are provided in \cref{numerics_details}

\begin{figure}
    \centering
    \includegraphics[width=1.0\linewidth]{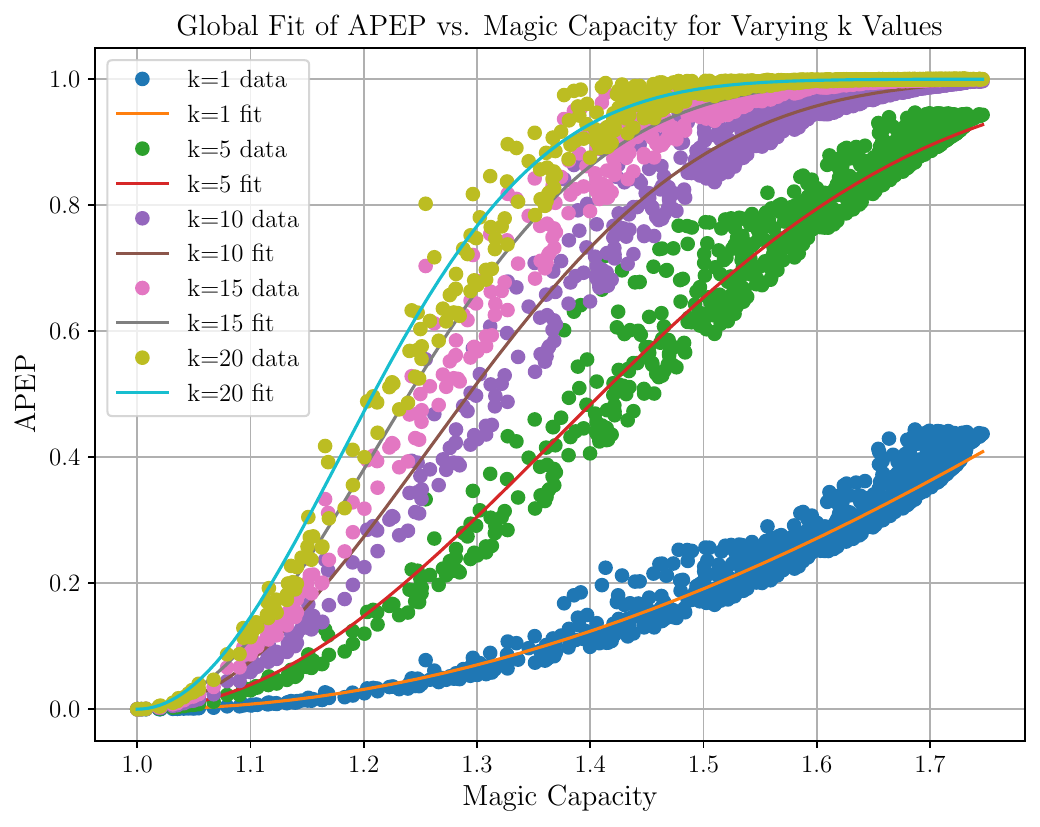}
    \caption{$\overline{P_E(\Omega_{C,\mathcal{U}})}^C$ vs. $\mathcal{K}(U)$ for select values of $k$.}
    \label{unitary_k_plot}
\end{figure}

We find that:

\begin{equation}
    \overline{P_E(\Omega_{C,\mathcal{U}})}^C = 1-\left\vert\cos(1.2567\cdot(\mathcal{K}(\Omega_{C,\mathcal{U}})-1))\right\vert^{k}.
\end{equation}

The data and fit demonstrate two important takeaways. First, while the relationship between APEP and magic capacity is not strictly monotonic, the best-fit curves all are. Generally, insertion of local gates with more magic yields more nonlocal magic. 

Second, the relationship is highly non-linear.  This is significant for designing circuits that generate nonlocal magic via application of local non-Clifford gates. In real quantum processing platforms, single-qubit gates often have a power and time cost associated with the Bloch sphere rotation angle $\theta$ \cite{theta_cost_1, theta_cost_2, theta_cost_3}. Each individual gate application has an associated error, and application of simultaneous single-qubit gates can induce correlated errors \cite{corr_err_1, corr_err_2, corr_err_3}. \Cref{unitary_k_plot} demonstrates that a given APEP value can be achieved by many different $\mathcal{K}$-$k$ pairs. Given the aforementioned costs and errors, there may be significant trade-offs and considerations in selecting a $\mathcal{K}$-$k$ ``scheme" for creating a nonlocal magic-generating circuit.

\section{Conclusion}\label{conclusion}

We have studied information scrambling and nonlocal magic generation in noisy Clifford encoding-decoding circuits. In the limit of large circuits ($L \rightarrow \infty$), our analytic work, relying on recent results in the representation theory of the Clifford group, yields simple formulae for the average values of the bipartite $\mathcal{A}$-OTOC and averaged Pauli-entangling power (APEP) over all Clifford encodings.

Like in \cite{butterfly_effect}, the values of these normalized quantities remain finite in the thermodynamic limit, displaying a butterfly effect in which noise on finitely many qubits yields macroscopic scrambling and nonlocal magic generation in an infinite system. To better elucidate these results, we explicitly work through the examples of depolarizing and unitary ($R_Z$) noise. We observe that for depolarizing noise, random Clifford encodings have both no magic-generating power, and no $\mathcal{A}$-OTOC due to decoherence effects overwhelming the information scrambling. For unitary noise, increasing the number of noisy qubits increases both the nonlocal magic generating power as well as the information scrambling of Clifford circuits.

We then study the relationship between an operator-level magic monotone for the single-qubit noise unitaries, called the \textit{magic capacity}, and the APEP. Interestingly, the APEP displays a dependence on the cosine of the magic capacity that scales exponentially in the number of qubits on which the magic unitary is applied. Since the APEP is a magic monotone, this relationship also holds for setups in which qubits just undergo magic gates followed by a random Clifford (without an initial ``encoding" Clifford). This means that a single-qubit unitary with a low magic capacity, if applied to many qubits, can generate as much nonlocal magic as a higher magic capacity unitary acting on fewer qubits, indicating a non-trivial trade-off when engineering such systems.

Further exploration of such a tradeoff in the context of realistic hardware constraints would extend these results to answer very pragmatic questions. Such literature exists in the context of preparing magic states \cite{magic_cost, msd_low, msd_factory, msd_opt} but not yet at the operator level. Of course, this would benefit from a deeper operational interpretation of the APEP in particular, and nonlocal magic in general. For instance, the precise relationship between nonlocal magic and simulability of noisy quantum circuits is unknown. 

Furthermore, the APEP is only meaningful for closed systems. It is therefore not useful for understanding how \textit{arbitrary} noise propagates magic and is not applicable to realistic dissipative systems. Extending an APEP-like quantity to open systems would dramatically broaden the scope of this work.

\section{Acknowledgements}\label{acknowledgements}

ED acknowledges valuable discussions with Faidon Andreadakis. ED and PZ acknowledge partial support from the NSF award PHY2310227.

\bibliographystyle{unsrt}
\bibliography{references}

\appendix\label{appendix}
\begin{widetext}

\section{Clifford Averages of the A-OTOC and APEP}\label{clifford_schurweyl}

Let the Kraus operators of $\mathcal{E}^{\otimes k} \otimes \mathbb{I}^{\otimes (L-k)}$ be $\{ \Lambda_{\alpha} \}_\alpha$. It was shown in \cite{butterfly_effect} that 

\begin{equation}\label{4designequation}
    G(\Omega_{C,\mathcal{E}})=\frac{1}{d^2}\Tr\left(T_{(13)(24)} (C^\dagger)^{\otimes 4} \left(\sum_\alpha \Lambda_\alpha \otimes \Lambda_\alpha^\dagger \right) C^{\otimes 4} \left(T_{(12)}^A\otimes(d_B T_{(34)} - T_{(34)}^A) \right) \right).
\end{equation}

To calculate $\overline{G(\mathcal{\Omega})}^C$, we must to evaluate the following (and plug back into the above formula): $$\mathbb{E}_C\left[\left(C^{\dagger}\right)^{\otimes 4} \left( \sum_\alpha \Lambda_\alpha \otimes \Lambda_\alpha^\dagger \right) C^{\otimes 4}\right].$$

Similarly, to calculate $\overline{P_E(\Omega_{C,\mathcal{E}})}^C$ \cref{APEP_eqn}, with coherent noise $\mathcal{E} = \mathcal{U} = U(\cdot)U^\dagger$, we must evaluate $$ \mathbb{E}_C\left[\left(C^{\dagger}\right)^{\otimes 4} ((U^\dagger)^{\otimes k} \otimes \mathbb{I}^{\otimes (L-k)}) Q (U^{\otimes k} \otimes \mathbb{I}^{\otimes (L-k)})  C^{\otimes 4}\right] $$.

\subsection{Schur-Weyl Duality}

Here, we follow the articulate notation of the supplementary material of \cite{Turkeshi}. To simplify the subsequent equations, let

\begin{equation}
    \Phi_{\text {Cliff }}^{(k)}(O)=\mathbb{E}_{C \in \mathcal{C}\left(2^L\right)}\left[\left(C^{\dagger}\right)^{\otimes k} O C^{\otimes k}\right] \equiv \int_{\text {Haar }} d \mu(C)\left(C^{\dagger}\right)^{\otimes k} O C^{\otimes k}.
\end{equation}\footnote{Note that this integral can be written simply as a uniform average over all the elements in the Clifford group, as the group is finite. We write it as an integral to maintain consistency with the circular unitary ensemble notation.}

Via Schur-Weyl duality \cite{weingarten_1} for the Clifford group \cite{cliff_comm}, we can write:
\begin{equation}
    \Phi_{\text {Cliff }}^{(k)}(O)=\sum_{\pi \in \mathcal{S}_k} \left[ b^+_\pi(O) QT_\pi + b^-_{\pi}(O)(\mathbb{I}^{\otimes 4}-Q)T_\pi \right],
\end{equation}

where $\mathcal{S}_k$ is the permutation group over $k$ elements, $T_\pi$ is the representation of the permutation $\pi$ acting on $\left(\mathbb{C}^{2^L}\right)^{\otimes k}$, and $Q \equiv \frac{1}{d^2}\sum_{P \in \tilde{\mathcal{P}}_N} P^{\otimes 4}$ as in the main text. Note $T_\pi=\otimes_{i=1}^N t_\pi^{(i)}$, where $t_\pi^{(i)}$ is the representation of $\pi$ on the $i$-th qubit; each permutation is the tensor of single-qubit permutations across the $k$ copies of the system Hilbert space.

For the Clifford group:
\begin{equation}\label{traces}
    b^+_\pi(O) = \sum_{\sigma \in S_k} W_{\pi, \sigma}^+ \Tr\left(OQ T_\sigma\right) \ ; \quad 
    b^-_\pi(O) = \sum_{\sigma \in S_k} W_{\pi, \sigma}^- \Tr\left(O(\mathbb{I}^{\otimes 4}-Q) T_\sigma\right)
\end{equation}

where $W_{\pi, \sigma}$ is the generalized \textit{Weingarten symbol} corresponding to the permutations $\pi, \sigma$. 

Recall that for the symmetric group $\mathcal{S}_k$, the irreps correspond to the integer partitions of $k$. An integer partition of $k$ is an (ordered, for uniqueness) set of positive integers which add up to $k$. Let $\lambda$ refer to an arbitrary integer partition of $k$, let $\Pi_\lambda$ be the projection onto the irrep (corresponding to) $\lambda$, $d_\lambda$ be the irrep's dimension, and $\chi_\lambda$ be its character function. Then
\begin{equation}
    W_{\pi, \sigma}^\pm=\sum_{\lambda \vdash k} \frac{d_\lambda^2}{(k!)^2} \frac{\chi_\lambda(\pi \sigma)}{D^\pm_\lambda},
\end{equation}

where $D^+_\lambda = \Tr(QT_\lambda)$ and $D^-_\lambda = \Tr((\mathbb{I}^{\otimes 4}-Q)T_\lambda)$.

To calculate the traces in \cref{traces}, we use the following identity. Let $T_\sigma \in \mathcal{S}_4$ have the cyclic representation $(ab)(cd)$, where $\left\{a...d\right\} \in \left\{1,2,3,4\right\}$\footnote{The choice of a $(2,2)$ cycle structure here is arbitrary and made to illustrate the identity.}. Let $M = \otimes_{k=1}^4 M_k$. Then

\begin{equation}
    \Tr(T_\sigma M) = \Tr(M_a M_b)\Tr(M_c M_d).
\end{equation}

This is a straightforward generalization of the above swap formula.

Finally, to obtain the formulae in \cref{mainresults}, we implemented the above formulae in a Mathematica notebook.

\section{APEP and Magic Capacity Numerics}\label{numerics_details}

We adapt the semi-definite programming code from \cite{capacity_robustness} to numerically determine the magic capacity for $1000$ randomly sampled single-qubit unitaries $U$, and we exactly calculate the corresponding $\overline{P_E(\Omega_{C,\mathcal{U}})}^C$, for $k$ values from $1$ to $20$ using \cref{apep_unitary_clifford}. Explicitly, each $\Omega_{C,\mathcal{U}} = \mathcal{C}^\dagger (\mathcal{U}^{\otimes k} \otimes \mathbb{I}^{\otimes (L-k)}) \mathcal{C}$ for a randomly selected Clifford $C$.

We jointly fit the data for all $20$ $k$ values via the Levenberg-Marquardt algorithm \cite{levenberg_marquardt}. We then bootstrap our dataset $1000$ times to reduce idiosyncratic sampling variability for, and to provide uncertainty estimates of, the ansatz parameters $a, \ b$. To four decimal places, this returned $a = 1.2567 \pm 0.0009$ and $b = 1.0000 \pm 0.0000$, with $95\%$ confidence intervals $\left[1.2551, 1.2584 \right]$ and $\left[1.0000, 1.0000 \right]$ respectively. \Cref{bootstrap_histograms} displays the distribution of different parameter values over the $1000$ resamplings.

\begin{figure*}[t!]
    \centering
    \begin{subfigure}[t]{0.48\textwidth}
    \includegraphics[width=0.8\linewidth]{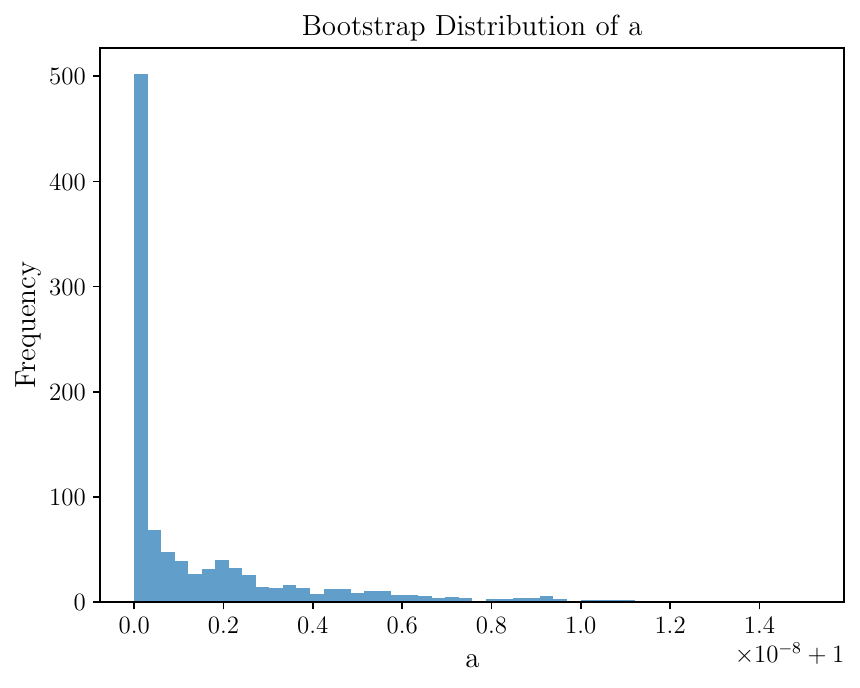}
    \caption{Bootstrap histogram of $a$ parameter values ($n = 1000$ resamplings).}
    \end{subfigure}%
    ~
    \begin{subfigure}[t]{0.48\textwidth}
    \includegraphics[width=0.8\linewidth]{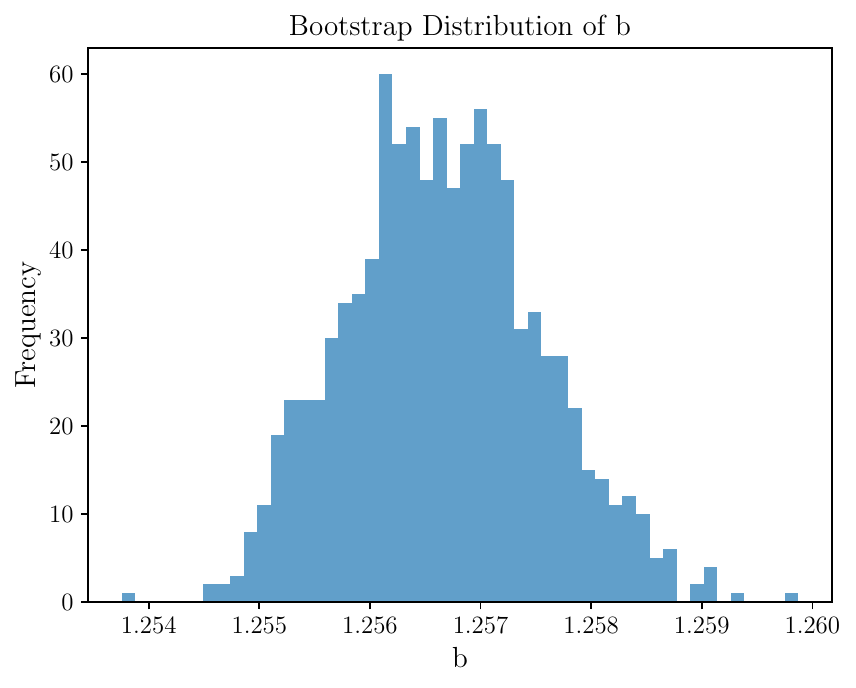}
    \caption{Bootstrap histogram of $b$ parameter values ($n = 1000$ resamplings).}
    \end{subfigure}
    \caption{Bootstrap histograms for $a$ and $b$ parameters in global fit of average APEP vs. magic capacity.}
    \label{bootstrap_histograms}
\end{figure*}

\section{Typicality of Clifford-Averaged Quantities}\label{typicality_numerics}

We are interested in the Clifford-averaged quantities in the main text to the extent that they capture ``typical" behavior of randomly selected individual Clifford encodings. Unlike for the full unitary group, however, there are no known measure concentration results for functions on the Clifford group under the unitary measure. To make an analytic claim about the closeness of the value of the bipartite $\mathcal{A}$-OTOC or APEP for a particular Clifford to the average would require the computation of the variance of the Clifford-averaged quantities in the main text. As these are functions of the fourth moment of the Cliffords, the variance is an eighth moment quantity; this is too computationally resource-intensive to implement.

We thus provide some numerics here, for small qubit values $L$, to display the decrease of the variance of both quantities as a function of $L$. This serves as preliminary evidence that variance decreases as $L \rightarrow \infty$.

To calculate the APEP for a \textit{particular} $\Omega_{C,\mathcal{U}}$, we employ \cref{APEP_eqn}. Recall that $\Omega_{C,\mathcal{U}} = \mathcal{C^\dagger}(\mathcal{U}^{\otimes k} \otimes \mathbb{I}^{\otimes (L-k)})\mathcal{C}$, but also that $P_E(\Omega_{C,\mathcal{U}}) = P_E(\Omega'_{C,\mathcal{U}})$, where $\Omega'_{C,\mathcal{U}} = \mathcal{C^\dagger}(\mathcal{U}^{\otimes k} \otimes \mathbb{I}^{\otimes (L-k)})$. Using the fact that the action of $T^A_{(12)(34)} $ commutes with partial tracing over $B$, along with the generalized swap trick, it can be shown that \begin{equation}
    P_E(\Omega'_{C,\mathcal{U}}) = 1-\frac{1}{d^2}\sum_{P \in \tilde{P}_N}\left(\Tr\left(\Tr_B(C^\dagger (U^{\otimes k} \otimes \mathbb{I}^{\otimes (L-k)}) P ((U^\dagger)^{\otimes k} \otimes \mathbb{I}^{\otimes (L-k)} ) C) \right)\right)^2.
\end{equation} 

As this requires only a single Hilbert space copy, this expression can be computed far more efficiently than \cref{APEP_eqn}. To analyze the behavior of the variance, we randomly sample $n_U = 4$ random single-qubit unitaries $U$, and $n_C = 12$ random Cliffords $C$ for each $L$, from $L = 3$ to $L = 8$. For each $L$ and $U$, we calculate $P_E(\Omega_{C,\mathcal{U}})$ for each $C$ and then compute the variance of the set of these values. We then take the mean of the variance over the randomly sampled $U$ and plot this average variance versus $L$. We do this for $U$ acting on $k = 1$ to $k = 3$ qubits, yielding $3$ curves. We observe in \cref{average_variances} that the average variance decreases with increasing $L$.

For the bipartite $\mathcal{A}$-OTOC, we use the following equivalent expression shown in \cite{NamitBipartite}:
\begin{proposition}\label{aotoc_state}
    We denote by $\psi:=|\psi\rangle\langle\psi|$ with $|\psi\rangle \in \mathcal{H}_A$. Then,
    \begin{equation}
        G(\Omega)=N_A \mathbb{E}_\psi\left[S_L\left(\operatorname{Tr}_B \widetilde{\Omega}(\psi)\right)-d_B\left(S_L(\widetilde{\Omega}(\psi))-S_L^{\min }\right)\right]
    \end{equation}
where $\mathbb{E}_\psi$ is the the Haar average over pure states on $\mathcal{H}_A, N_A:=\frac{d_A+1}{d_A}$, the linear entropy $S_L(\rho) \equiv 1 - \Tr(\rho^2)$, $S_L^{\min }:=1-\frac{1}{d_B}$, and $\tilde{\Omega}_{C,\mathcal{U}}(X) \equiv \Omega_{C,\mathcal{U}}(X \otimes \frac{\mathbb{I}}{d_B})$. 
\end{proposition}

With this, we estimate the bipartite $\mathcal{A}$-OTOC for a given $\Omega_{C,\mathcal{U}}$ (i.e. choice of single-qubit noise $U$ and Clifford $C$) by sampling $n_\psi   = 8$ pure states on $A$ and averaged their resultant \cref{aotoc_state} expressions. We do this for $n_C = 50$ Cliffords and then take the variance of this set, then average the variance over $n_V = 10$ single-qubit noise unitaries. We run this simulation for $k=1$ to $k=3$ noisy sites and $L=4$ to $L=8$ qubits. The average variance is then plotted against $L$. Like the APEP, we observe that the average variance decreases with increasing $L$, and the variance of this average variance over different $U$ also decreases. This offers some initial evidence that randomly selected Clifford encodings undergoing arbitrary coherent noise cluster around the Clifford-averaged value of the bipartite $\mathcal{A}$-OTOC.

\begin{figure*}[t!]
    \centering
    \begin{subfigure}[t]{0.48\textwidth}
    \includegraphics[width=0.8\linewidth]{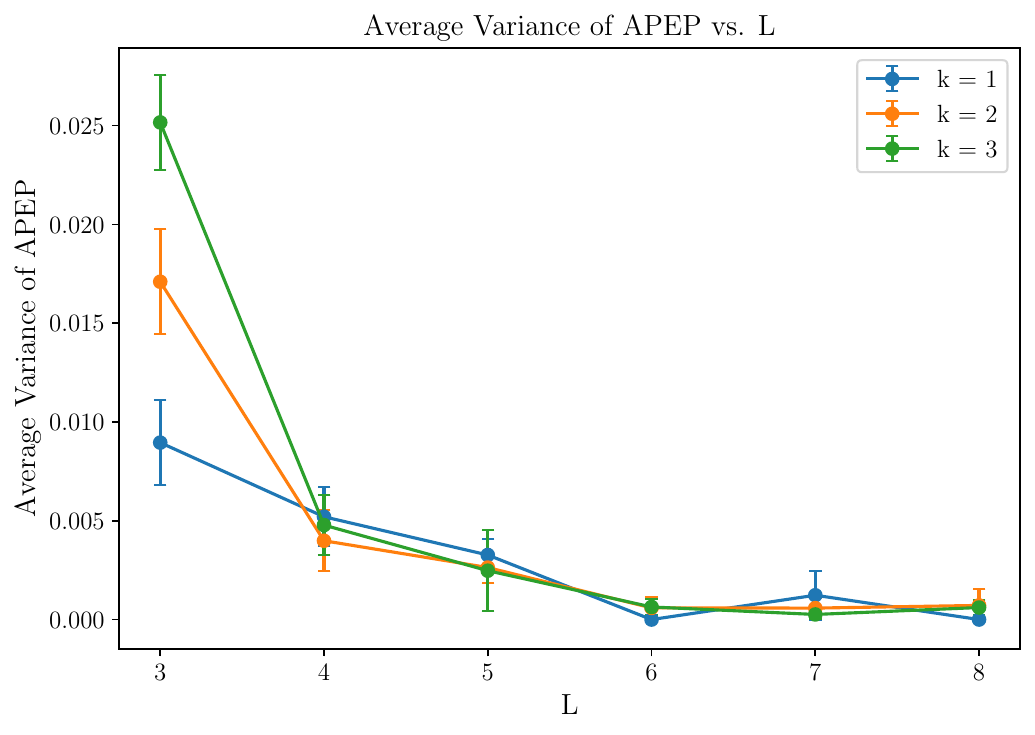}
    \caption{Average variance of APEP vs. $L$ for $L = 3$ to $L = 8$, plotted for $U$ acting on $k = 1$ up to $k=3$ qubits.}
    \end{subfigure}%
    ~
    \begin{subfigure}[t]{0.48\textwidth}
    \includegraphics[width=0.8\linewidth]{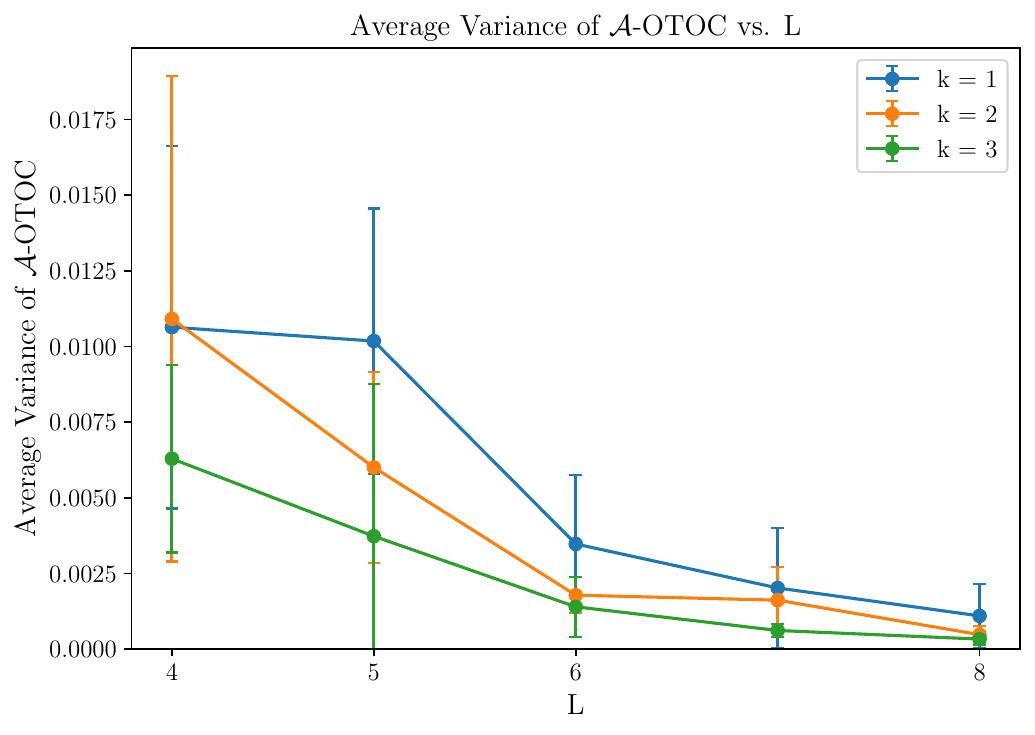}
    \caption{Average variance of $\mathcal{A}$-OTOC vs. $L$ for $L = 4$ to $L = 8$, plotted for $U$ acting on $k = 1$ up to $k=3$ qubits.}
    \end{subfigure}
    \caption{Variances of APEP and $\mathcal{A}$-OTOC vs. $L$ for $k = 1$ to $k = 3$. For both the APEP and the bipartite $\mathcal{A}$-OTOC, the plotted error bars represent the standard error of the average variance over different single-qubit coherent noise $U$.}
    \label{average_variances}
\end{figure*}

\end{widetext}

\end{document}